**Choosing a growth curve to model the Covid-19 outbreak**

Emiro A. Molina-Cuevas

Departamento de Matemáticas Puras y Aplicadas,
Universidad Simón Bolívar, Caracas, Venezuela.
Gerencia de Estadísticas,
Corporación Multifranquicias, C.A., Caracas, Venezuela.
*emolina@usb.ve*

**Abstract**

The Richards family models comprise a well-known set of models with useful parameters to describe several aspects of disease outbreaks. Some of these models have been used to study the current Covid-19 pandemic. However, there seems to be confusion regarding the discrimination among competing models. In this paper a detailed application of Akaike's information approach is used to discern among models using data from The European Union, The United States and The United Kingdom. We argue that the epidemiological characteristics derived from competing models should be examined to complement the selection strategy, and the implicit properties of the models contrasted with the available data. Detailed analytical expressions of the epidemiological characteristics implied by the selected parametrizations are also offered.

**Key Words**: Covid-19, Richards models, Gompertz models, Logistic models, AIC, Parametrization, Growth Rate, Inflection point,

**1. Introduction**

The Coronavirus Covid-19 pandemic is the subject of intense research in many areas of the scientific community. Consequently, a plethora of research papers addressing the subject of the dynamics of the spread of the disease in the population has been published. The traditional approach accommodates several strategies. Growth curve models (e.g. the Logistic, Bertalanffy and Gomperzt growth curves), differential equations prediction models (such as the widely studied SIR and SEIR models), time series models, stochastic models, machine learning strategies and many more. See, *e.g.*, Roosa, Lee, Luo, *et al.* (2020); Jia, Li, Jiang, *et al.* (2020); Jung, Akhmetzhanov, Hayashi, *et al.* (2020); Ardabili, Mosavi, Ghamisi, et al. (2020). The models are used to generate insights into the transmission dynamics of the disease, assess the potential impact of different intervention strategies and perform short-term forecasts. In the light of the lockdown rules imposed in several countries to reduce the Covid-19 rate of contagions and the urgency to reopen their economies, the forecasting of the number of afflicted, deaths and recoveries, is of particular interest.

In this paper we focus on the use of a family of traditional parametric growth curve models, the Richards family, to examine the cumulative number of daily confirmed cases. These models are based on basic assumptions on the rate of growth and the expected behaviour of the diseases. The curves fitted here are sigmoids and the model parameters provide important characteristics of the growth of the disease: the upper asymptotic value (the upper bound for the number of cases, deaths or recoveries), the inflection point (the turning point where the growth rate starts to decrease), the time at the inflection point, the maximum relative growth rate and the lower asymptote, when present. Notable members of this family in epidemiological literature are the Bertalanffy, Logistic, Generalized Logistic and Stannard growth curves. Richards (1959) introduced his model as a generalization of Verhulst's (1838) three parametric logistic model and

von Bertalanffy's (1938) growth function in the context of ecological population growth. A three-parameter version of the Richards model, the Chapman–Richards model, was proposed by Chapman (1961), in the context of the dynamics of fisheries populations. It can be shown that another well stablished model in the literature, the Gomperzt (1825) growth curve, is a limit case of one of the Richards parametrizations. However, as Tjørve and Tjørve (2010) highlight, "there is … considerable confusion in the literature because models are frequently presented in many forms (different parameterizations and re-parameterizations), and sometimes the same model is given different names, perhaps because the author(s) simply did not recognize the model". Recent examples of the application of these models to the study of Covid-19 are, *e.g*., Jia, Li, Jiang, *et al*. (2020) and Roosa, Lee, Luo, *et al*. (2020). An excellent review of the Richards family of models in the epidemiological context can be found in Tjørve and Tjørve (2010). For a connection between the Richards and SIR models see Wang, Wu and Yang (2012). For a detailed look at the mathematical properties of the curves the reader is referred to Panik (2014, Chap. 3).

An additional issue is the strategy for model selection. Although it should be well known that $R^2$ (adjusted or unadjusted) is not well suited as a measure for model selection on its own (Kvalseth, 1985, Spiess and Neumeyer, 2010), this measure is still being used to discriminate between competing models (see, *e.g*., Jia, Li, Jiang, *et al*. 2020). The measure is particularly inadequate to discriminate among non-linear models. In an extensive simulation study, Spiess and Neumeyer (2010) compared the performance of the adjusted measure to discriminate among several sigmoid models. They found that the adjusted $R^2$ was rarely affected more than in the third or fourth decimal, rendering it useless for the task. That does not mean that $R^2$ is inadequate as a measure of goodness of fit, particularly in the case of linear models. In that context $R^2$ reveals the proportion of the data that is explained by the fit. This interpretation cannot be transferred to the non-linear case since the total sum of squares is not equal to the regression sum of squares plus the residual sum of squares. Our analysis agrees with these observations. All models confronted here fit well into the range of observed data. Correspondingly the values of the adjusted $R^2$ are remarkably similar, differing only from the third or fourth decimal onwards. Several more appropriate measures for discrimination between models have been proposed in the literature (See, *e.g*., McQuarrie and Tsai, 1998). In this work we will use the Akaike (1974) family of measurements to compare models.

The models imply several epidemiological characteristics that should be examined to decide their consistency with the observed outbreak. In this paper we examine the carrying capacity, time of inflection, growth rate and predicted length of the outbreak. The analytical expressions for these characteristics are given explicitly for each of the proposed parametrizations, and the difference between the models along them is discussed. Other aspects implied by the models are also confronted with the reported behaviour of the outbreak.

## 2. Methods

### 2.1 Data

We used the public COVID-19 data at the GitHub website of the Center for Systems Science and Engineering (2020), Whiting School of Engineering, Johns Hopkins University. The data reports the number of accumulated confirmed cases, accumulated deaths, accumulated recovered cases and accumulated active cases per country; for the United States, the data includes each state. The data for each day is stored in a separate file and is updated daily. We consolidated the daily data of the countries in the European Union as a block and the daily data for the entire United States. The models were adjusted for the accumulated daily confirmed cases of the United States, the European Union, and the United Kingdom. The analysis was carried out using the data from the beginning of the series up to April 28, 2020.

2.2 Models

Let $Y(t)$ denote the number of accumulated confirmed cases of the pandemic at time $t$. The following models were considered.

*Richards model:*

Richards (1959) derived his model as an extended form of von Bertalanffy's (1938) growth function to plant data. Many parametrizations of the Richards model have been proposed (Tjørve and Tjørve, 2010). In this paper we use the version

$$Y(t) = a/\left(1 + e^{-b(t-\tau)}\right)^{1/d} \qquad (2.1)$$

This parametrization was first discussed, we believe, by Schenute (1981), who referred to it as a Bertalanffy's form of the model. Here the parameter $a$ represents the upper asymptote of the curve (the limiting value as $t$ tends to infinity, also referred to as the carrying capacity of the model), $b$ is a growth-rate parameter (which controls the slope at an inflection), $\tau$ is a location parameter and $d$ is a shape parameter that allows the model to have a floating inflection point. The inflection occurs at the point $t = t_{\text{inf}}$ for which the second derivative $Y''(t) = 0$ and is given by:

$$t_{inf} = \frac{b\tau - \log(d)}{b}, \qquad (2.2)$$

so that $Y(t_{\text{inf}}) = a/(1+d)^{1/d}$, a fraction of the upper asymptote which depends only on $d$. This parametrization has the $x$-axis as a lower asymptote. A three-parametric version of the model, the Chapman-Richards model (Chapman, 1961) is obtained by taking the parameter $\tau$ to be zero.

*Gomperzt model:*

The Gompertz (1825) curve was originally designed to describe age distribution in human populations. We tested both, a four parametric and a three parametric version of the Gomperzt growth model. The three parametric model is given by

$$Y(t) = ae^{-e^{-b(t-\tau)}} \qquad (2.3)$$

Here $a$ is the upper asymptote of the curve, $b$ sets the growth rate and $\tau$ is a location parameter with the property that $t_{\text{inf}} = \tau$, the inflection point. It follows that $Y(t_{\text{inf}}) = a/e$, where $e$ denotes Euler's number Therefore $Y(t_{\text{inf}})$ is at approximately 0.3679 times the value of the asymptote. Thus, the placement of the inflection point is not as flexible as with the Richards model. A slight improvement can be obtained by adding a fourth parameter:

$$Y(t) = c + ae^{-e^{-b(t-\tau)}} \qquad (2.4)$$

With this parametrization the inflection still occurs at $t_{\text{inf}} = \tau$. However now $Y(t_{\text{inf}}) = a/e + c$ and the asymptote is displaced to $a+c$.

*The Logistic growth model:*

Proposed by Verhulst (1838), the three parametric logistic model is perhaps the best known and most widely used growth model in epidemiology. We use the common parametrization

$$Y(t) = a/(1 + e^{-b(t-\tau)}) \tag{2.5}$$

This is a sigmoid with the lower asymptote at zero and the upper asymptote at *a*. The parameter *b* controls the rate of growth and *τ* is a location parameter

It follows that $Y''(t) = 0$ for $Y(t_{inf}) = a/2$ (half the value of the upper asymptote). Then a substitution of this Y value into Equation 2.5 yields $t_{inf} = \tau$, where it is assumed that $a > 1$ so that $\log(a) > 0$. Hence, the logistic function has a point of inflection at $(\tau, a/2)$, and is symmetric around this point. This property might be too restrictive for some natural growth processes. As with the Gompertz model the addition of a fourth parameter helps:

$$Y(t) = c + a/(1 + e^{-b(t-\tau)}). \tag{2.6}$$

With this parametrization $t_{inf} = \tau$, however, the asymptote and value at the point of inflection are displaced *c* units.

The three parametric version 2.5 performed poorly with the data at hand and will not be discussed further. Only the fourth parametric Logistic model 2.6 will be considered.

**3. Results**

3.1 Fitting the models

We fitted the models using the *nl* procedure of the statistical software Stata, version 16.1 (StataCorp, 2019b, p.p. 1704-1715). The program uses least squares to maximize the log-likelihood of the models and fit the parameters.

To compare the goodness of the models, the corrected Aikaike's (1974) Information Criterion (AICc), which penalizes models with a greater number of parameters, was used. Let $L(\hat{\theta}/y)$ denote the likelihood of a particular model where $\hat{\theta}$ denotes the estimated parameters and *y* the observed data. Let *p* denote the number of parameters in the model. Then the Aikaike's Information Criterion measure is

$$\text{AIC} = -2 \log(L(\hat{\theta}/y)) + 2p$$

and the corrected AIC is

$$\text{AICc} = \text{AIC} + p(p+1)/(n-p+1).$$

Suppose there are *k* competing models. Let $\text{AICc}_i$ denote the AICc for model *i* and let $\text{AICc}_{min}$ denote the AICc for the model with the minimum AICc among the *k* models. The Akaike difference for model *i* relative to the model with the minimum AICc is $\Delta_i = \text{AICc}_i - \text{AICc}_{min}$. The larger the $\Delta_i$ is, the less plausible it is that model *i* is the best model among the candidates. The best model among the set has $\Delta_i = \Delta_{min} = 0$.

The likelihood of model *i* given the data is proportional to $exp(-½\Delta_i)$ (Burnham and Anderson, 2002, p.p. 75-79). These model likelihoods can be normalized so that they sum to 1, giving rise to the Akaike weights:

$$w_i = \frac{exp\left(-\frac{\Delta_i}{2}\right)}{\sum_{j=1}^{k} exp\left(-\frac{\Delta_j}{2}\right)}.$$

The Akaike weight $w_i$ can be interpreted as the probability that model $i$ is the best model given the data and the set of models considered. The smaller the weight $w_i$, the less plausible is model $i$. The Akaike's weights can also be used to determine the relative importance of the models among the set.

The Evidence Ratio of model $j$ relative to model $i$ is defined as

$$\frac{w_j}{w_i} = \frac{exp\left(-\frac{\Delta_j}{2}\right)}{exp\left(-\frac{\Delta_i}{2}\right)}.$$

consequently, the Evidence Ratio of the best model with respect to model $i$ is $w_{min}/w_i = exp\left(\frac{\Delta_i}{2}\right)$. Note that for every model in the set these are the odds in favour of the model with minimum AICc with respect to the competing model. The higher these ratios, the lower the chances of the competing model against the best model. That is, the greater the evidence *against* the competing model.

Table 1 shows the AICc, $\Delta_i$, Akaike Weights and Evidence ratios (of the best model with respect to the competing model) corresponding to the Richards, 4-parametric and 3-parametric Gompertz and the 4 parametric Logistic model when fitted to the European Union, United States and United Kingdom data.

The Akaike's statistics are revealing. For the European Union data, the Richards model has a 0.935 probability of being the best model among the set given the actual data, which is a remarkably high probability. The importance given by the weights to the rest of the models is low compared to the importance given to the Richards model. The odds of the Richards model being the best are nearly 18 times greater than those of the 4-parametric Gompertz model, and 74 times greater than those of the 3-parametric Gompertz model. The odds of the Logistic model being the best model are extremely low (3.23E+31 times lower than the odds of the Richards model).

It is noteworthy to observe the corresponding adjusted $R^2$ statistics. Since all the models fit relatively well within the range of the data (Figure 1), the values are high, with differences from the third or fourth decimal on. Up to the fourth decimal the Richards and three-parametric Gompertz models share the same adjusted $R^2$ value, rendering the statistic useless to discriminate between the two models. Using the adjusted $R^2$ on its own as a ranking criterion, the three-parametric Gompertz model would rank better than the four-parametric Gompertz model thanks to a slight difference in the fourth decimal. This difference, on the other hand, is meaningless since the models are not nested. By contrast, the Akaike's Evidence Ratio between the two Gomperzt models tells us that the four-parametric version is 0.052/0.013 = 4 times more plausible than the three-parametric version with this data.

For the United States data, the best model under the Akaike's criterion is the 3-parametric Gompertz model. However, the evidence in favour of this model is not strong. The evidence ratios against the 4-parametric Gompertz and the Richards models are small. The probability of the model being the best alternative among the set is below the 0.5 level, and the chances of the

**Table 1.**
Akaike's model selection statistics between the best performing model and the other candidate models for the European Union, the United States, and the United Kingdom data sets.

| Regions | Model Statistics | Richards | Gompertz (4 parametric) | Gompertzt3 | Logistic (4 parametric) |
|---|---|---|---|---|---|
| European Union | AICc | 1571.038 | 1576.807 | 1579.657 | 1716.142 |
| | $\Delta_i$ | 0.000* | 5.770 | 8.620 | 145.104 |
| | Akaike's Weights | 0.935 | 0.052 | 0.013 | 2.90E-32 |
| | Evidence ratio | 1.000 | 17.902 | 74.425 | 3.23E+31 |
| | Adjusted $R^2$ | 0.9999 | 0.9998 | 0.9999 | 0.9988 |
| United States | AICc | 1635.389 | 1635.243 | 1634.069 | 1788.021 |
| | $\Delta_i$ | 1.320 | 1.174 | 0.000* | 153.9518 |
| | Akaike's Weights | 0.249 | 0.268 | 0.482 | 1.79E-34 |
| | Evidence ratio | 1.935 | 1.799 | 1.000 | 2.69E+33 |
| | Adjusted $R^2$ | 0.9998 | 0.9997 | 0.9998 | 0.9977 |
| United Kingdom | AICc | 1130.717 | 1132.122 | 1133.577 | 1267.17 |
| | $\Delta_i$ | 0.000* | 1.405 | 2.860 | 136.4534 |
| | Akaike's Weights | 0.576 | 0.286 | 0.138 | 1.35E-30 |
| | Evidence ratio | 1.000 | 2.019 | 4.179 | 4.27E+29 |
| | Adjusted $R^2$ | 0.9999 | 0.9999 | 0.9999 | 0.9990 |

*Best model according to Akaike's Information Criterion*

Richards model being the best are nearly 25%. The only clear-cut difference is with respect to the Logistic model, which remains, by far, the last choice among the competing models under all the criteria. The fitted Richards model for these data has $\tau = 0$, so strictly speaking, this is the three-parametric version of the Richards model.

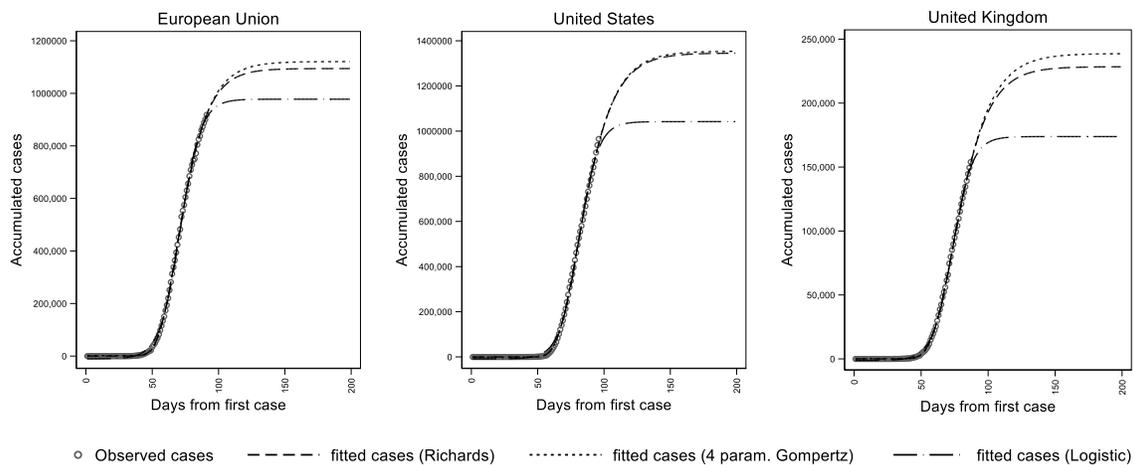

**Figure 1**.
Observed vs. fitted accumulated cases by region. The 3-parametric Gomperzt fit has been omitted since the plot is too close to the 4-parametric plot to be distinguished.

For the United Kingdom data, the Akaike's Criterion favours the Richards model, with a probability of being the best model of 0.576. The odds against the two Gompertz models are not drastic, although the chances of the 3-parametric Gompertz model are less than 14%. Again, the evidence against the logistic model is overwhelming. It is worth noting that, up to the fourth decimal, the adjusted $R^2$ is the same for all four models in the case of the United Kingdom.

The Logistic model can be safely ruled out from the competing models for all three data sets. Regarding the Richards and Gomperzt models, for the European Union data the Richards model is clearly the most plausible model, however, for the United Kingdom and the United States the support for the best model is not as strong. We highlight that the Richards model is the best or as good as the best model in all cases. However, as pointed out by Burnham and Anderson (2002, page 80), inference must admit that there are sometimes competing models and the data do not support selecting only one.

### 3.2 Beyond Goodness of fit

Judging goodness of fit is only one step in selecting a model (or models) for the data at hand. In our case the selected model must have epidemiological meaning and be consistent with the properties of the observed data. The fitted parameters for each model are listed in table 2. From

**Table 2.**
Estimated Model Parameters for the European Union, the United States, and the United Kingdom data sets.

| Regions | Model Parameters | Richards | Gompertz (4 parametric) | Gompertz (3 parametric) | Logistic (4 parametric) |
|---|---|---|---|---|---|
| European Union | a | 1094301.1*** (10044.9) | 1118818.9*** (6290.2) | 1124431.5*** (6502.6) | 988318.8*** (9175.8) |
| | b | 0.0782*** (0.0023) | 0.0718*** (0.0007) | 0.0713*** (0.0007) | 0.1300*** (0.0025) |
| | τ | 40.75*** (4.832) | 68.53*** (0.107) | 68.52*** (0.107) | 71.20*** (0.184) |
| | c | - | 2017.5*** (441.9) | - | -10471.2*** (2488.2) |
| | d | 0.110** (0.0340) | - | - | - |
| United States | a | 1345755.6*** (17312.3) | 1354948.9*** (50869.7) | 1350034.8*** (17560.9) | 1050410.7*** (18859.7) |
| | b | 0.0660*** (0.0010) | 0.0653*** (0.0024) | 0.0656*** (0.0010) | 0.1370*** (0.0036) |
| | τ | 0 (fixed) | 80.33*** (0.621) | 80.30*** (0.232) | 80.98*** (0.323) |
| | c | - | -889.6* (372.0) | - | -8403.8** (2522.3) |
| | d | 0.0050*** (0.0003) | - | - | - |
| United Kingdom | a | 228462.1*** (5330.8) | 238535.8*** (3769.0) | 240191.1*** (2615.6) | 175059.6*** (2442.6) |
| | b | 0.0673*** (0.0029) | 0.0617*** (0.0010) | 0.0612*** (0.0007) | 0.1360*** (0.0026) |
| | τ | 35.660*** (8.274) | 73.810*** (0.305) | 73.890*** (0.201) | 73.470*** (0.241) |
| | c | - | 205.0*** (45.9) | - | -1213.0*** (307.5) |
| | d | 0.0774* (0.0353) | - | - | - |

Standard errors in parentheses; * $p < 0.05$, ** $p < 0.01$, *** $p < 0.001$

these parameters, several epidemiological characteristics might be derived. In table 3 we show the following characteristics: *The Upper Horizontal Asymptote*, which sets the upper threshold level for the number of cases attainable under the model; the *Inflection Point*, the number of days from day one at which the rate of growth changes from increasing to decreasing; the predicted *Length of the Outbreak* in days from day one, and the *Maximum Rate of Growth*, which occurs at the inflection point, denoted by $r_{max}$ in the table. The standard errors and confidence intervals reported in the table were obtained with Stata's procedure *nlcom* (StataCorp, 2019b, p.p. 1235 - 1243). The program implements the δ-method to derive the standard errors; an asymptotic gaussian distribution is assumed to obtain confidence intervals.

**Table 3.**
Estimated Model Characteristics for the European Union, the United States, and the United Kingdom data sets.

| Regions | Models | Characteristics | Estimates | S.E. | [95% Conf. Interval] Lower limit | [95% Conf. Interval] Upper limit |
|---|---|---|---|---|---|---|
| European Union | Richards | Upper Asymptote | 1094301 | 10045 | 1074291 | 1114312 |
| | | Inflection point | 68.9232 | 0.1513 | 68.6266 | 69.2198 |
| | | Length of outbreak | 246.6519 | 5.1480 | 236.5620 | 256.7418 |
| | | $r_{max}$ | 0.0705 | 0.0005 | 0.0695 | 0.0715 |
| | Gompertz (4 parametric) | Upper Asymptote | 1120836 | 6444 | 1108207 | 1133466 |
| | | Inflection point | 68.5263 | 0.1032 | 68.3239 | 68.7286 |
| | | Length of outbreak | 262.3737 | 2.1076 | 258.2429 | 266.5045 |
| | | $r_{max}$ | 0.0715 | 0.0004 | 0.0707 | 0.0723 |
| | Gompertz (3 parametric) | Upper Asymptote | 1124431 | 6503 | 1111481 | 1137382 |
| | | Inflection point | 68.5228 | 0.1070 | 68.3097 | 68.7359 |
| | | Length of outbreak | 264.7367 | 2.0017 | 260.8136 | 268.6599 |
| | | $r_{max}$ | 0.0713 | 0.0007 | 0.0699 | 0.0727 |
| | Logistic (4 parametric) | Upper Asymptote | 977848 | 8030 | 962109 | 993587 |
| | | Inflection point | 71.2045 | 0.1842 | 70.8375 | 71.5714 |
| | | Length of outbreak | 177.3765 | 2.2011 | 173.0624 | 181.6906 |
| | | $r_{max}$ | 0.0664 | 0.0008 | 0.0648 | 0.0681 |
| United States | Richards | Upper Asymptote | 1345756 | 17312 | 1311289 | 1380222 |
| | | Inflection point | 80.2878 | 0.2324 | 79.8323 | 80.7434 |
| | | Length of outbreak | 294.1308 | 3.5115 | 287.2484 | 301.0132 |
| | | $r_{max}$ | 0.0657 | 0.0003 | 0.0650 | 0.0663 |
| | Gompertz (4 parametric) | Upper Asymptote | 1354059 | 50567 | 1254949 | 1453169 |
| | | Inflection point | 80.3302 | 0.6207 | 79.1137 | 81.5468 |
| | | Length of outbreak | 296.6393 | 9.1733 | 278.6600 | 314.6186 |
| | | $r_{max}$ | 0.0654 | 0.0005 | 0.0645 | 0.0663 |
| | Gompertz (3 parametric) | Upper Asymptote | 1350035 | 17561 | 1315074 | 1384996 |
| | | Inflection point | 80.2995 | 0.2325 | 79.8367 | 80.7623 |
| | | Length of outbreak | 295.5634 | 3.6033 | 288.5012 | 302.6257 |
| | | $r_{max}$ | 0.0656 | 0.0010 | 0.0636 | 0.0675 |
| | Logistic (4 parametric) | Upper Asymptote | 1042007 | 17880 | 1006964 | 1077050 |
| | | Inflection point | 80.9764 | 0.3225 | 80.3342 | 81.6186 |
| | | Length of outbreak | 182.5040 | 3.0702 | 176.4864 | 188.5215 |
| | | $r_{max}$ | 0.0694 | 0.0010 | 0.0674 | 0.0713 |
| United Kingdom | Richards | Upper Asymptote | 228462 | 5331 | 2178245 | 239100 |
| | | Inflection point | 73.6608 | 0.1984 | 73.2719 | 74.0497 |
| | | Length of outbreak | 256.8986 | 8.3700 | 240.4937 | 273.3035 |
| | | $r_{max}$ | 0.0625 | 0.0003 | 0.0619 | 0.0631 |
| | Gompertz (4 parametric) | Upper Asymptote | 238741 | 2631 | 233584 | 243897 |
| | | Inflection point | 73.8125 | 0.1984 | 73.4166 | 74.2085 |
| | | Length of outbreak | 274.6347 | 2.6640 | 269.4134 | 279.8560 |
| | | $r_{max}$ | 0.0615 | 0.0002 | 0.0611 | 0.0619 |
| | Gompertz (3 parametric) | Upper Asymptote | 240191 | 2616 | 234973 | 245409 |
| | | Inflection point | 73.8937 | 0.2008 | 73.4931 | 74.2943 |
| | | Length of outbreak | 276.4462 | 2.5780 | 271.3935 | 281.4989 |
| | | $r_{max}$ | 0.0612 | 0.0007 | 0.0598 | 0.0625 |
| | Logistic (4 parametric) | Upper Asymptote | 173847 | 2310 | 169320 | 178374 |
| | | Inflection point | 73.4735 | 0.2411 | 72.9923 | 73.9546 |
| | | Length of outbreak | 162.4835 | 1.9706 | 158.6213 | 166.3458 |
| | | $r_{max}$ | 0.0688 | 0.0006 | 0.0675 | 0.0700 |

For the Richards and three-parametric Gompertz models the upper horizontal asymptote is given by the parameter $a$. For the rest of the models the value corresponds to the sum $a + c$.

As discussed earlier, the inflection point for the Richards model is given by (2.2), and it is equal to the parameter $\tau$ for the rest of the models. To determine the length of the outbreak we seek for the point $t_m$ such that $Y(t_m) = Asymptote - 1$. This value is attainable by the function $Y(t)$ and its integer part can be associated with a real date. Alternatively, a percentage of the asymptotic value could be used. However, we prefer this method, since when evaluated at integer values (days) this is the last integer value that the function can take.

For the Richards model the value is given by:

$$t_m = \frac{\tau b - \log\left[\left(\frac{a-1}{a}\right)^{-d} - 1\right]}{b}.$$

For the four and three parametric Gompertz curves the value is:

$$t_m = \tau - \left\{\frac{\log\left[\log\left(\frac{a}{a-1}\right)\right]}{b}\right\}.$$

For the four parametric Logistic model the values is:

$$t_m = \tau + \frac{\log(a-1)}{b}.$$

The instantaneous growth rate at time *t* is defined as (see *e.g.* Panik, 2014, Chap. 3)

$$r(t) = Y'(t)/Y(t). \qquad (3.1)$$

It has its maximum value $r_{max}$, the *maximum growth rate*, at the inflection point, $r_{max} = r(t_{inf})$. For the Richards model, $r_{max} = b/(d+1)$; for the 4 parametric Gompertz, $r_{max} = ab/(a+ce)$, where *e* denotes Eulers's number; for the three parametric Gompertz model $r_{max}$ reduces to *b*; for the four parametric logistic model, $r_{max} = ab/(2a+4c)$.

Since the estimates are the means of asymptotically normal random variables, a Wald Chi-square test can be used to compare the differences between the estimated characteristics reported in table 3 for each model in each region. As an illustration, table 4 presents the values of this test and its significance for the differences between the characteristics associated with the Richards model and those associated with the other three models, and the differences between the characteristics associated with the four-parametric Gompertz model and the characteristics associated with the three-parametric Gompertz and four-parametric Logistic models.

For the United States, there are no significant differences between the epidemiological characteristics derived from the Richards and Gompertz models. However, the differences are highly significant with the characteristics predicted from the Logistic model. We recall that the Akaike statistics favoured the three-parametric Gompertz model; however, the evidence as compared to the Richards and four-parametric Gompertz models were not strong. Thus, the use of any of these three models would render similar results. The Logistic model's carrying capacity (upper asymptote) and predicted length of outbreak seem low for the observed behaviour; the model clearly underperforms with this data when compared to the other models. For the European Union data the observed differences between the characteristics associated with the Richards model and those associated with the rest of the models are significant, with the exception of the differences in the maximum growth rate of the Gompertz models. There are no significant differences between the characteristics derived from the two Gompertz models. The differences of the characteristics for the Logistic model and the rest of the models are highly significant. Again, the carrying capacity and predicted length of outbreak derived from the Logistic model seem low for the observed behaviour. Akaike's strategy strongly suggests choosing the Richards model for the European Union data; this will lead to conclusions that will differ significantly from those obtained by using the other models. For the United Kingdom data, the conclusions from the Akaike's strategy favoured the Richards model, although the comparison with the four-parametric Gompertz model was not as clear cut as in the case of the European Union data. The carrying capacity and time of inflection of these models do not differ significantly when using the United Kingdom data, but the maximum growth rate and predicted length of outbreak do. The two models will render similar conclusions if chosen. As with the other two data sets, the Logistic model should not be recommended.

Table 4.
Wald Chi-square values for the differences between the characteristics reported in table 3.

| Regions | Characteristics | Wald tests for differences between Richards characteristic and the characteristics of the rest of the models | | | Differences between the 4 param. Gompertz characteristics and the | |
|---|---|---|---|---|---|---|
| | | Gompertz (4 parametric) | Gompertz (3 parametric) | Logistic (4 parametric) | Gompertz (3 parametric) | Logistic (4 parametric) |
| European Union | Upper Asymptote | 4.9436* | 6.3399* | 81.9994*** | 0.1542 | 192.8717*** |
| | Inflection point | 4.6965* | 4.6685* | 91.5911*** | 0.0006 | 160.8967*** |
| | Length of outbreak | 7.9878** | 10.7202** | 153.0966*** | 0.6609 | 777.9331*** |
| | $r_{max}$ | 2.4390 | 0.8649 | 18.8876*** | 0.0615 | 32.5125*** |
| United States | Upper Asymptote | 0.0241 | 0.0301 | 148.9562*** | 0.0057 | 33.8499*** |
| | Inflection point | 0.0041 | 0.0013 | 3.0008 | 0.0021 | 0.8535 |
| | Length of outbreak | 0.0652 | 0.0811 | 572.7205*** | 0.0119 | 139.2123*** |
| | $r_{max}$ | 0.2647 | 0.0092 | 12.5596*** | 0.0320 | 12.8000*** |
| United Kingdom | Upper Asymptote | 2.9896 | 3.9012* | 88.3644*** | 0.1527 | 343.5423*** |
| | Inflection point | 0.2923 | 0.6807 | 0.3598 | 0.0827 | 1.1788 |
| | Length of outbreak | 4.0772* | 4.9817* | 120.5598*** | 0.2388 | 1145.5107*** |
| | $r_{max}$ | 7.6923** | 2.9138 | 88.2000*** | 0.1698 | 133.2250*** |

$^{*} p < 0.05$, $^{**} p < 0.01$, $^{***} p < 0.001$

If the models are consistent with the observed behaviour of the outbreak, the new daily cases reported in the data at day $t$ should correlate closely with the consecutive daily differences $D(t) = Y(t) - Y(t-1)$ obtained from the models, $t = 2, 3, \ldots$. The estimated differences for each model and the observed daily new cases are plotted in figure 2 for each region. The maximum of the fitted curves corresponds to the Inflection points of the respective $Y(t)$ models. The plots do not suggest any inconsistencies between the fitted and observed curves. Table 4 shows the correlation between the estimated and observed new daily cases and the slopes resulting from

**Table 5.**
Slopes of regressions through the origin of fitted differences $D(t)$ on observed new daily cases and the corresponding correlation coefficients.

| Regions | Coefficients | Models | | | |
|---|---|---|---|---|---|
| | | Richards | Gompertz (4 parametric) | Gompertz (3 parametric) | Logistic (4 parametric) |
| European Union | Slope | 1.005*** (0.028) | 1.002*** (0.028) | 1.002*** (0.028) | 1.005*** (0.032) |
| | Correlation | 0.886 | 0.886 | 0.886 | 0.849 |
| United States | Slope | 1.020*** (0.016) | 1.019*** (0.015) | 1.020*** (0.016) | 1.019*** (0.027) |
| | Correlation | 0.968 | 0.968 | 0.968 | 0.905 |
| United Kingdom | Slope | 1.006*** (0.023) | 1.004*** (0.023) | 1.004*** (0.023) | 1.015*** (0.028) |
| | Correlation | 0.934 | 0.934 | 0.934 | 0.904 |

Standard errors in parentheses; $^{*} p < 0.05$, $^{**} p < 0.01$, $^{***} p < 0.001$

fitting a regression through the origin of the fitted on the observed values. As can be seen, the slopes are close to the ideal value of one. Although the correlations are smaller for the Logistic model in all the regions, the differences among the slopes in each region are not significant. These simple checks show no evidence of inconsistencies between the fitted models and the observed data.

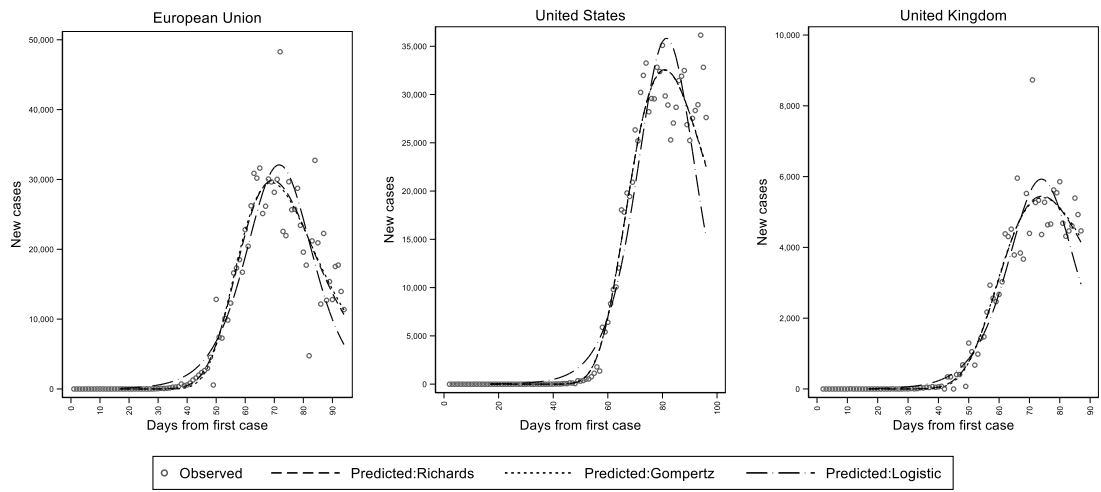

**Figure 2**.
Observed vs. fitted new daily cases by region. The 3-parametric Gomperzt fit has been omitted since the plot is too close to the 4-parametric plot to be distinguished.

*A caveat*
The models considered are static *cetaris paribus* concoctions. The models assume a "regular" behaviour under fixed conditions. However, to reduce the rate of contagion governments have imposed several "lockdown" strategies. These interventions lead to different rates of disease growth and while these are in place one might expect the curves to follow a particular growth pattern. However, these strategies are difficult to implement and come at a high social and economic cost. Thus, the modalities of these strategies differ greatly, and lifting the interventions is a priority. This might drastically change the behaviour of the disease growth. The pattern might change to the point of inducing a new inflection in the curves, a feature that none of the discussed models can handle. However, the models are particularly useful to assess the early stages of the diseases. As the famous quote of G.E.P. Box (1976) affirms, "*All models are wrong but some models are useful*".

**4. Conclusions**

The four and three parametric versions of the Richards and Gompertz models outperformed the logistic model for the analysed cases. Under the information criterion approach the Richards model outperformed the other models or performed as well as the best model, a similar result to that of Tjørve and Tjørve (2010). This is mainly due to the flexibility of the inflection point's position in relation to the asymptote. For the opposite reason (the symmetry of the curve around the inflection point) the Logistic model tends to descend too soon after the range of the observed data, into its asymptotic behaviour. This results in a predicted carrying capacity significantly lower than that predicted by the Richards and Gompertz models. There is a strong consistency among the epidemiological characteristics implied by the Richards and Gompertz models, and a clear-cut difference could not always be established with the available data. Still, the flexibility showed by the Richards model makes it a great contender when modelling the early stages of the growth of the Covid-19 outbreak.